\documentclass{revtex4}

\usepackage{amsmath}
\usepackage{amssymb}
\usepackage{amsthm} 

\usepackage{array}

\usepackage{slashed}

\usepackage[dvips]{graphicx} 
\usepackage{subfigure}

\newcommand\nn{\nonumber}
\newcommand\ba{\begin{eqnarray}}
\newcommand\ea{\end{eqnarray}}
\newcommand\eq[1] {\begin{align} #1 \end{align}}   
\newcommand\ga[1] {\begin{gather} #1 \end{gather}}   

\newcommand{\br}[1]{ \left( #1 \right) }
\newcommand{\brs}[1]{ \left[ #1 \right] }
\newcommand{\brf}[1]{ \left\{ #1 \right\} }
\newcommand{\brm}[1]{\left| #1 \right|}

\newcommand{\Sp}{\mbox{Sp}}
\newcommand{\Tr}{\mbox{Sp}}
\renewcommand{\Re}{\mbox{Re}}

\newcommand{\vv}[1]{{\bf #1}}

\newcommand{\dd}[1]{{\slashed #1}}   

\newcommand{\GeV}{\mbox{GeV}}
\newcommand{\MeV}{\mbox{MeV}}

\newcommand{\M} {{\cal M}} 
\newcommand{\A} {{\cal A}} 

\begin{document}

\title{On the measurement of $\chi_2(^3P_2)$ quarkonium state in the processes $e^++e^-\to \bar p+p$ and $\bar p+p\to e^++e^-$}

\author{E.A.~Kuraev}
\email{kuraev@theor.jinr.ru}
\affiliation{\it JINR-BLTP, 141980 Dubna, Moscow region, Russian Federation}

\author{Yu.M.~Bystritskiy}
\email{bystr@theor.jinr.ru}
\affiliation{\it JINR-BLTP, 141980 Dubna, Moscow region, Russian Federation}

\author{V.V.~Bytev}
\email{bvv@jinr.ru}
\affiliation{\it JINR-BLTP, 141980 Dubna, Moscow region, Russian Federation}

\author{E.~Tomasi-Gustafsson}
\email{etomasi@cea.fr}
\affiliation{\it CEA,IRFU,SPhN, Saclay, 91191 Gif-sur-Yvette Cedex, France, and \\
CNRS/IN2P3, Institut de Physique Nucl\'eaire, UMR 8608, 91405 Orsay, France}

\date{\today}

\begin{abstract}
The intermediate quarkonium state $\chi_2(^3P_2)$ in electron-positron
annihilation to proton and antiproton
as well as in antiproton-proton annihilation to electron and positron can produce backward-forward asymmetry, when populated through two photon exchange.
We use the dispersion relation method, which permits to express the asymmetry in terms of
partial widths of quarkonium decay.
The asymmetry dependence on the center of mass energy in the range near the resonance is presented.
The comparison with a similar effect in these reactions with the neutral $Z$-boson in the intermediate state is given. We show that these effects are $\le 10^{-3}$. The main source of asymmetry is of pure QED origin ($\sim 10^{-2}$) which arises from the interference between initial and final state real photon emission.
\end{abstract}

\maketitle

\section{Introduction}

\begin{figure}
    \centering
    \mbox{
        \subfigure[]{\includegraphics[width=0.3\textwidth]{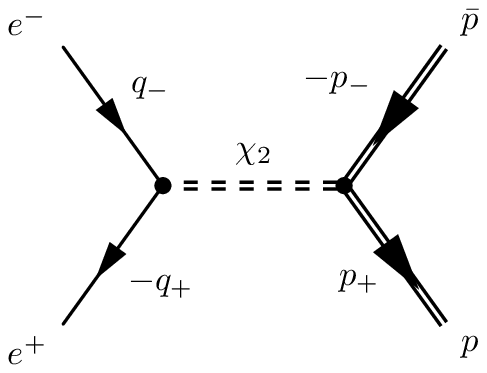}\label{fig.EEChiPP}}
        \quad\quad
        \subfigure[]{\includegraphics[width=0.3\textwidth]{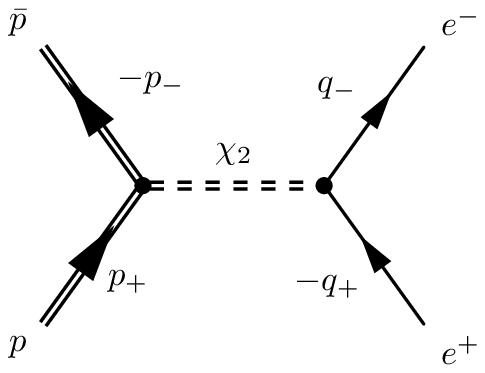}\label{fig.PPChiEE}}
    }
    \caption{Feynman diagrams of processes $e^+ +e^-\to \bar p+p$ and $\bar p+p\to e^++e^-$
    with quarkonium intermediate states.}
    \label{fig.QuarkoniumIntermediateStates}
\end{figure}

In a recent paper \cite{Zhou:2011yz} it was proposed that a large C-odd effect could arise in the electron-positron
annihilation into a hadron pair through two photon exchange which couple to resonances such as
$\Phi= \eta$, $\eta_C$, $\chi_0$, $\chi_2$, $\rho$, $a_1$.
In one loop level, taking into account the two-photon intermediate state
(Figs.~\ref{fig.EEChiPP} and \ref{fig.PPChiEE}),
there
is the possibility to access the $^1S_0$, $^3P_0$, $^3P_2$ quarkonium intermediate states,
which are bound states of charm quark and antiquark.
It was shown in Refs. \cite{Zhou:2011yz,Ebert:2003mu} that the contribution to the asymmetry arising
from the intermediate states $^1S_0$,  $^3P_0$ is proportional to the ratio of the electron mass to the
proton mass and it is negligible.
The intermediate state with quantum numbers $^3P_2(3556)=\chi_2$ can be produced by two virtual photons,
which arise from the annihilation subprocess of
electron and positron.
In Ref. \cite{Zhou:2011yz} this contribution to asymmetry was calculated and it was indicated that it is
responsible for a
large asymmetry, which can reach up to $40\%$. In this case, two photon exchange would be experimentally
observable in present and planned experiments.

In principle two photon exchange is suppressed by the factor $\alpha=1/137$, the electromagnetic fine structure
constant. Therefore, specific mechanisms should be advocated, which can compensate such suppression.

In this note we recalculate the excitation of the $\chi_{c2}$ resonance, through two photon exchange, and
show that the asymmetry in this case does not exceed level of $10^{-3}$. We compare this value with the
asymmetry generated by Z-boson exchange, which we find also of the order of
$10^{-3}$,  and by the asymmetry of pure QED nature, due to the interference
of photon emission in initial and final state \cite{Ahmadov:2010ak}, which turns out to be the most
important source of asymmetry ($\sim 10^{-2}$).

\section{Born approximation}

At the facilities VEPP (Novosibirsk), BEPC (Beijing), and PANDA/FAIR (Darmstadt), one can measure the
process of creation of proton and anti-proton in electron-positron annihilation:
\eq{
e^+(q_+)+e^-(q_-)\to \bar p(p_-)+p(p_+)
\label{eq:eqBES}
}
and the process of annihilation of proton and antiproton into electron and positron
\eq{
\bar p(p_-)+p(p_+)\to e^+(q_+)+e^-(q_-)
\label{eq:eqPANDA}
}
in the energy range of the mass of the quarkonia bound states of charmed quark and anti-quark.
\begin{figure}
    \centering
    \mbox{
        \subfigure[]{\includegraphics[width=0.3\textwidth]{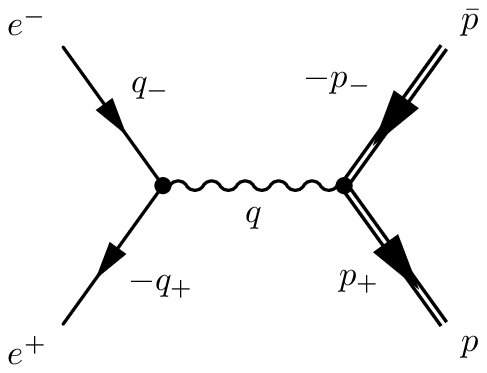}\label{fig.EEgPP}}
        \quad\quad
        \subfigure[]{\includegraphics[width=0.3\textwidth]{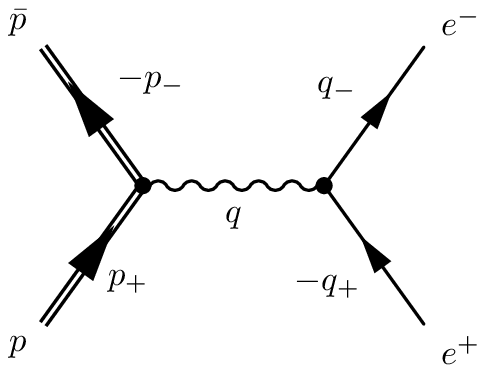}\label{fig.PPgEE}}
    }
    \caption{Feynman diagrams of processes $e^+ +e^-\to \bar p+p$ and $\bar p+p\to e^++e^-$
    in Born approximation.}
    \label{fig.BornApproximation}
\end{figure}
In Born approximation the amplitude of these processes has the form (see Figs.~\ref{fig.EEgPP} and \ref{fig.PPgEE}):
\eq{
\M_{\gamma}&= \frac{4\pi\alpha}{s} \brs{\bar v(q_+) \gamma^\mu u(q_-)} \brs{\bar u(p_+)\Gamma_{\mu}(q) v(p_-)},
\label{eq.Born}
}
where $q=p_++p_-=q_++q_-$, $s = q^2$ and electromagnetic vertex of the proton $\Gamma_{\mu}(q)$ is
parametrized in terms of
two form factors:
\eq{
\Gamma_\mu(q) &=\gamma _\mu F_1(q^2)+\frac{1}{4M_p} \br{ \dd{q} \gamma_\mu-\gamma_\mu\dd{q}} F_2(q^2),
}
where $F_1$ and $F_2$ are the Dirac and Pauli form factors of the proton in the time-like region of
momentum transfer ($q^2 > 4M_p^2$) which
related with the Sachs electric and magnetic proton form factors as
\eq{
G_E=F_1+\frac{s}{4M_p^2}F_2, \qquad G_M=F_1+F_2.
}
The square modulus of the matrix element (\ref{eq.Born}) has the form
\eq{
    \sum_{spins} |\M_\gamma|^2=4(4\pi\alpha)^2\brs{|G_M|^2 (1+c^2)+(1-\beta^2) |G_E|^2 (1-c^2)},
    \qquad
    \beta=\sqrt{1-\frac{4M_p^2}{s}},
}
where $\beta$ is the proton velocity and $c \equiv \cos\theta$, $\theta$ is the angle between vectors
$\vv{q_-}$ and $\vv{p_+}$ in the center of mass system of the initial particles.
The cross sections of processes (\ref{eq:eqBES}) and (\ref{eq:eqPANDA}) in Born approximation
thus has the form:
\eq{
\frac{d\sigma^{p\bar p\to e^+e^-}}{dc}=\frac{1}{8s\beta} \frac{1}{16\pi}\sum_{spins}|\M_\gamma|^2, \qquad
\frac{d\sigma^{e^+e^-\to p\bar p }}{dc}=\frac{1}{8s} \frac{\beta}{16\pi}\sum_{spins}|\M_\gamma|^2. \label{eq:eqcs}
}
The interference of the Born amplitude $\M_\gamma$ from (\ref{eq.Born}) with a  single photon in the intermediate
state with the amplitude with quarkonium in the intermediate
state (see Figs.~\ref{fig.EEChiPP} and \ref{fig.PPChiEE}) originates backward-forward asymmetry
\eq{
\A = \frac{d\sigma(c)- d\sigma(-c)}{d\sigma(c)+d\sigma(-c)},
\label{eq:Asymmetry}
}
where $d\sigma(c)=d\sigma/dc$ is the angular dependent cross section.
Let us estimate this asymmetry.

\section{Asymmetry from $\chi_2$ : approximation of partial widths}

The amplitude of the processes (\ref{eq:eqBES}) and (\ref{eq:eqPANDA})
with the $\chi_{c2}$ intermediate state corresponds to the
diagrams on Figs.~\ref{fig.EEChiPP} and \ref{fig.PPChiEE} and can be written as:
\eq{
\M_{\chi_2}
=
\frac{\M_{\chi_{c2}\to e^+e^-} \, \,\M_{\chi_{c2}\to p\bar p}}
{s-M_{\chi_2}^2+i M_{\chi_2}\Gamma_{\chi_2}},
\label{eq.PartialWidthsAmplitude}
}
where we used the "partial width" approximation which is acceptable within the
$\chi_{c2}$--resonance intermediate state vicinity.The matrix elements $\M_{\chi_{c2}\to e^+e^-}$ and $\M_{\chi_{c2}\to p\bar p}$ describe the
conversion of $\chi_2$ tensor particle into the electron--positron and proton--antiproton pairs correspondingly.
\begin{figure}
    \includegraphics[width=0.2\textwidth]{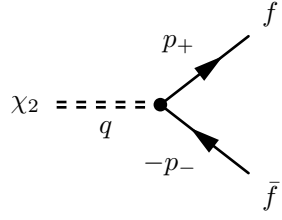}
    \caption{Feynman diagrams of the conversion of $\chi_2$ tensor particle into fermion--antifermion pair.}
    \label{fig.ChiToFF}
\end{figure}
This type of matrix elements can be parametrized in the following way \cite{Kuhn:1979bb} (see Fig.~\ref{fig.ChiToFF}):
\eq{
\M_{\chi_2 \to f\bar{f}}=g_f \brs{\bar{v}(p_+)\gamma^\mu u(p_-)} \br{p_+-p_-}^\nu\chi_{\mu\nu},
}
where $g_f$ is the corresponding coupling constant and $\chi_{\mu\nu}$ is the polarization tensor of $J=2$ state,
which has the following properties:
\eq{
\chi_{\mu\nu}&=\chi_{\nu\mu}, \qquad
\chi_{\mu\nu}g^{\mu\nu}=0, \qquad
\chi_{\mu\nu}q^{\mu}=0,\nn\\
\chi_{\mu\nu}\chi_{\lambda\sigma}&=\frac{1}{2}\br{ \Pi_{\mu\lambda}\Pi_{\nu\sigma} + \Pi_{\mu\sigma}\Pi_{\nu\lambda} }
- \frac{1}{3}\Pi_{\mu\nu}\Pi_{\lambda\sigma},
\label{eq:eqchi}\\
\Pi_{\alpha\beta}&=-g_{\alpha\beta}+  \frac{q_{\alpha}q_{\beta}}{ q^2}.
\nn
}
The decay width which corresponds to the amplitude $\M_{\chi_2 \to f\bar{f}}$ has the form:
\eq{
\Gamma_{\chi_2 \to f\bar{f}}=\frac{g_f^2 \,M_{\chi_2}^3}{120\pi}\beta_f^3(5-2\beta_f^2),
\qquad \beta_f=\sqrt{1-\frac{4m_f^2}{M_{\chi_2}^2}},
}
where $\beta_f$ is the velocity of final fermion $f$.
In particular
\eq{
\Gamma_{\chi_2 \to p\bar{p}}=\frac{g_p^2 \, M_{\chi_2}^3}{120\pi}\beta^3(5-2\beta^2),
\qquad
\Gamma_{\chi_2 \to e\bar{e}}=\frac{g_e^2 \, M_{\chi_2}^3}{40\pi}.
}
Keeping in mind possible complexity of the product $g_e g_p = \brm{g_e g_p} e^{i\phi}$,
we can write the interference of Born amplitude $\M_\gamma$ (see (\ref{eq.Born}))
with the amplitude $\M_{\chi_2}$ from (\ref{eq.PartialWidthsAmplitude}) which gives the following result:
\eq{
|\M|^2_{\rm int}
=
2\sum_{\rm spins} \Re\brs{\M_\gamma^*\M_{\chi_2}}
=
\frac{32\pi\alpha M_{\chi_2}^3}{\Gamma_{\chi}} \brm{g_e g_p}\frac{y\cos\phi+\sin\phi}{y^2+1}\beta c(1-\beta^2c^2),
}
where $y$ is the convenient variable which vanishes at the resonance region:
\eq{
    y=\frac{s-M^2_{\chi_2}}{M_{\chi_2}\Gamma_{\chi_2}}.
\label{eq:eqy}
}
Thus we can estimate the asymmetry $\A_\chi$ from (\ref{eq:Asymmetry}) as
\eq{
\A_\chi=\A_0\frac{y\cos\phi+\sin\phi}{y^2+1}\frac{\beta c(1-\beta^2c^2)}{2-\beta^2(1-c^2)},
\qquad
\A_0=\frac{\brm{g_e g_p} M_{\chi_2}^3}{\pi\alpha\Gamma_{\chi_2}},
}
where we used point-like proton approximation ($F_1=1$, $F_2=0$).

Unfortunately the partial decay width of the decay of $\chi_2$ state into the electron--positron pair
as well as the phase $\phi$ is not known.
Below we will calculate these phases in approximation of two photon and two gluon intermediate
state mechanisms, for the conversion of $\chi_2$ into electron-positron pair and into proton-anti-proton pair.

\section{Two photon and two gluon intermediate state approach}

Our approach for evaluating the conversion amplitudes consists in the calculation of their $s$-channel discontinuities and in the restoration of the
whole amplitude by means of dispersion relations. In this way we can express the results in terms of the experimentally measurable
partial width of conversion of $\chi_2$ tensor meson state to two photons and to hadrons.

We use the partial width approximation for the total amplitude $\M_\chi$ again (\ref{eq.PartialWidthsAmplitude}) and calculate the imaginary part as:
\eq{
    \Delta \brm{\M}^2_{\rm int}
    =
    2 \sum_{\rm spins} \Re \brs{\M_\gamma^* \Delta\M_{\chi_2}}.
}
The amplitudes
$\M_{\chi_{c2}\to e^+e^-}$ and $\M_{\chi_{c2}\to p\bar p}$ are estimated in the approximation of two photon and two gluon intermediate
state mechanisms:
\eq{
    \M_{\chi_{c2}\to e^+e^-} &= 4 (4\pi\alpha)^2 g_\gamma \chi_{\alpha\beta} T_{\alpha\beta\lambda}
    \nn\\
    \M_{\chi_{c2}\to p\bar p} &= 4 (4\pi\alpha_s) g_{g}c_{col} \chi_{\mu\nu} R_{\mu\nu\lambda}
%
%
\label{eq:eq18}}
where the dimensional constants $g_\gamma,g_g$ describe the conversion  of $^3P_2$ state to two photons and two gluons.
In Eq. (\ref{eq:eq18}) we include the color factor $I c_{col}=3(1/2)\delta^{ab}t^at^b=2I$, where $I$ is the unit matrix in the color space, which describes the interaction
of two gluons with each of the quark of the proton.
Moreover we define:
\eq{
T_{\alpha\beta\lambda}&=\frac{1}{4}Sp \hat{q}_-\gamma_\lambda\hat{q}_+O^e_{\alpha\beta}, \nn \\
O^e_{\alpha\beta}&=\int d\Phi_2\frac{1}{2q_-k_1}K_{\alpha\beta\rho\sigma}\gamma_\rho(\hat{q}_--\hat{k}_1)\gamma_\sigma; \nn \\
R_{\mu\nu\lambda}&=\frac{1}{4}Sp (\hat{p}_--M_p)\Gamma_\lambda(\hat{p}_--M_p)O^p_{\mu\nu}, \nn \\
R^p_{\mu\nu}&=\int d\Phi_2\frac{1}{2p_+k_1}K_{\mu\nu\rho\sigma}\gamma_\rho(\hat{p}_+-\hat{k}_1+M_p)\gamma_\sigma,
}
with
\eq{
K_{\alpha\beta\rho\sigma}=(k_1k_2)g_{\alpha\sigma}g_{\beta\rho}+k_{1\alpha}k_{2\beta}g_{\rho\sigma}-k_{1\alpha}k_{2\sigma}g_{\rho\beta}-k_{1\rho}k_{2\beta}g_{\alpha\sigma}.
}

The corresponding matrix elements and the partial widths are
\eq{
M_{2\gamma}&=g_\gamma K_{\mu\nu\alpha\beta}\chi_{\alpha\beta}e_1^\mu(k_1)e_1^\nu(k_2), \nn \\
\Gamma_{\chi_2\to \gamma\gamma}&=\frac{g_\gamma^2 M_\chi^3}{320\pi}; \nn \\
M_{gg}&=g_g \frac{1}{2}\delta^{ab}K_{\mu\nu\alpha\beta}\chi_{\alpha\beta}e_1^{a\mu}(k_1)e_1^{b\nu(k_2)}, \nn \\
\Gamma_{\chi_2\to gg}&=\frac{g_g^2 M_\chi^3}{160\pi}.
}
Keeping in mind the duality principle, we assume that the $\chi_2\to hadrons$ decay goes through two gluons at the leading order of $\alpha_s$. Thus we can identify $\Gamma_{\chi_2\to 2 gluons}$ with the hadronic width of $^3P_2$ state
\eq{
\Gamma_{\chi_2\to gg}=0.8\Gamma_{total}=1.6~\MeV,
\Gamma_{\chi_2\to \gamma \gamma}=0.2\Gamma_{total}=0.4~\MeV,
}
where the factor 0.8(0.2) takes into account the proportion of the hadronic(photonic) decay channels of $\chi_2$ \cite{Beringer:1900zz}.
This gives
\eq{
    g_\gamma = 3.3857 \cdot 10^{-4}~\GeV^{-1},
    \qquad
    g_g = 0.13~\GeV^{-1}.
}
The phase factor $e^{i\phi}$ is defined as $e^{i\phi}=e^{i(\phi_e+\phi_p)}$ where $e^{i\phi_e}$ and $e^{i\phi_p}$ arise respectively from the amplitudes $\M_{\chi_c2}\to e^+e^-$ and $\M_{\chi_c2}\to p\bar p$.
\subsection{Two photon intermediate state to the vertex $\chi_2\to e\bar{e}$}
For discontinuity of amplitude of conversion of the tensor state to the lepton pair through two-photon intermediate state $T$ we obtain
\eq{
\Delta_s T_{\alpha\beta\lambda}=4\pi\alpha g_\gamma\int d\Phi_2\frac{1}{2\br{k_1q_-}}K_{\alpha\beta\sigma\rho}\frac{1}{4}\Sp\brs{\dd{q_-}\gamma_\lambda\dd{q_+}
\gamma_\rho\br{\dd{q_-}-\dd{k_1}}\gamma_\sigma}.
}
Performing the integration on two photon volume
\eq{
 d\Phi_2=\int\frac{d^3k_1 d^3 k_2}{2\omega_1 2\omega_2(2\pi)^2}\delta^4(q-k_1-k_2),
}
we obtain (see details in Appendix~\ref{appendix.A})
\eq{
T_{\alpha\beta\lambda}=\frac{5\alpha}{72\pi}g_\gamma\br{\ln\frac{M_{\chi_2}^2}{m_e^2}-i\pi}\frac{1}{4}\Sp\brs{\dd{q_-}\gamma_\lambda\dd{q_+}\gamma_\alpha} (q_+-q_-)_\beta.
}
We can conclude that the phase associated with conversion to two photons $\phi_e$ is small
\eq{
tg\phi_e=\frac{\pi}{\ln\frac{M_{\chi_2}^2}{m_e^2}}=0.17.
}
So we assume further the conversion amplitude of $^3P_2$ state to the lepton pair to be real (i.e., $\phi_e \approx 0$).

\subsection{Total discontinuity, real part and asymmetry}

\begin{figure}
    \includegraphics[width=0.6\textwidth]{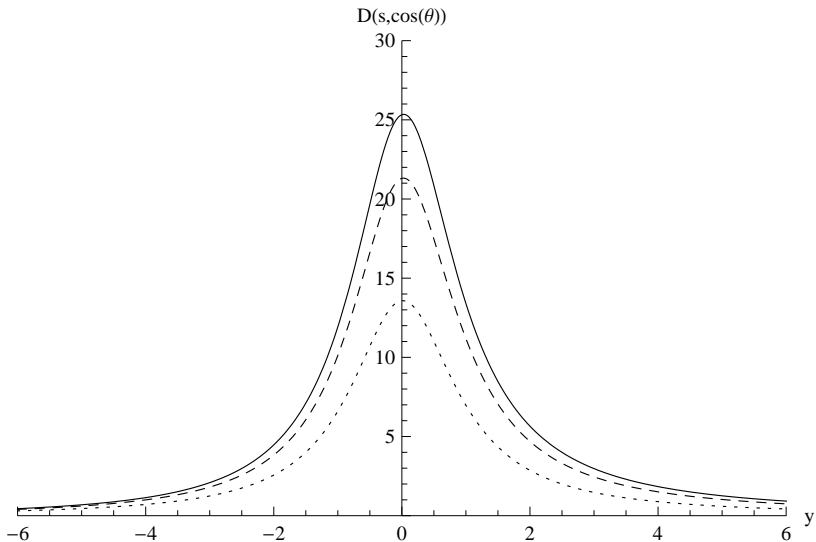}
    \caption{\label{fig.DEnergyPlot}
        The numerical estimation of the quantity $D(y,\cos(\theta))$ (see (\ref{eq.D})) as a function of $y$ within the region corresponding to $M_\chi-3\Gamma_\chi \leq \sqrt{s} \leq M_\chi+3\Gamma_\chi$.
        The solid line corresponds to $\theta=10^o$, dashed line --- $\theta=30^o$, dotted line --- $\theta=50^o$.
    }
\end{figure}

\begin{figure}
    \includegraphics[width=0.6\textwidth]{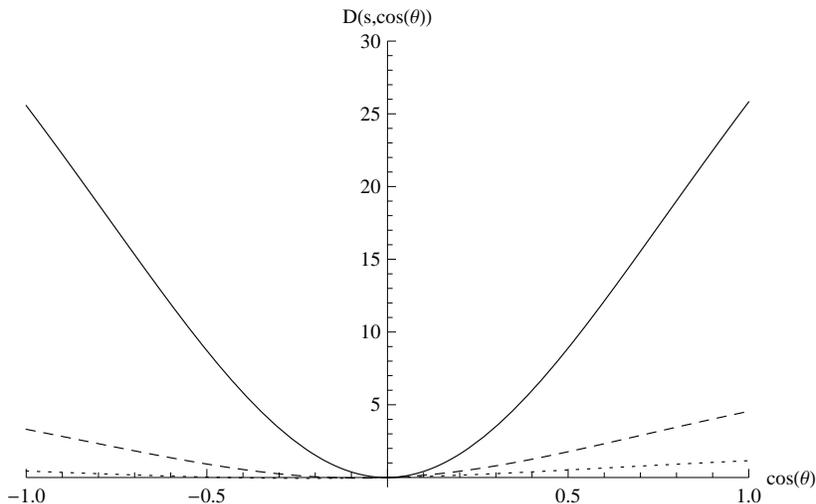}
    \caption{\label{fig.DAngularPlot}
        The numerical estimation of the quantity $D(y,\cos(\theta))$ (see (\ref{eq.D})) as a function of $c=\cos\theta$ for point-like proton (i.e. $F_1=1$, $F_2=0$).
        The solid line corresponds to $\sqrt{s}=M_\chi$, dashed line --- $\sqrt{s}=M_\chi-\Gamma_\chi$, dotted line --- $\sqrt{s}=M_\chi-2\Gamma_\chi$.
    }
\end{figure}

After conversion of tensor structures, omitting the phase associated with lepton part of the amplitude
and performing the angular integration (see Appendix~\ref{appendix.B}) we obtain for the total discontinuity
(we are interested only in the odd part of the contribution)
\eq{
    \brm{\M}_{int}^2&=
    \frac{10\pi}{27}\alpha^2\alpha_s(M_{\chi_2}^2g_\gamma g_g)\ln\frac{M_{\chi_2}^2}{m_e^2}\frac{s}{(s-M_{\chi_2}^2)^2+M_{\chi_2}^2\Gamma^2}
    \brs{
        \br{s-M_{\chi_2}^2} H_{odd}+M_{\chi_2}\Gamma G_{odd}
    }, \nn \\
    G_{odd}&=
    \frac{c}{\beta^3}\brs{-34\beta^4-6\beta^3+54\beta^2-18+9\frac{(1-\beta^2)^3}{\beta}L}+\nn \\
    &+c^2 \brs{54\beta^4+10\beta^3-6\beta^2+30+\frac{3}{\beta}\br{4\beta^5-6\beta^3+7\beta-5}L}, \qquad L=\ln\frac{1+\beta}{1-\beta},
}
where we used the approximation of point-like proton (i.e. $F_1 = 1$ and $F_2 = 0$) which is realistic near the threshold.
The real part, corresponding to $G_{odd}$, can be found using the dispersion relation (see also comments in Appendix~\ref{appendix.B}):
\eq{
    H_{odd}(\beta,c)=\frac{1}{2\pi}\brf{ G_{odd}(\beta,c)\ln\frac{1-\beta^2}{\beta^2}+\int\limits_0^1\frac{2x d x}{x^2-\beta^2}\br{-G_{odd}(\beta,c)+G_{odd}(x,c)}}.
\label{eq.DispersionRelation}
}
Setting $F_1=1$, $F_2=0$, we obtain for the asymmetry
\eq{
\A_\chi&=\A_0 D, \qquad D=\frac{y H_{odd}+G_{odd}}{y^2+1}\frac{1}{2-\beta^2(1-c^2)}, \label{eq.D} \\
\A_0&=\frac{5\alpha_s}{864\pi}(M^2g_\gamma g_g)\ln\frac{M_{\chi_2}^2}{m_e^2} = 1.86 \cdot 10^{-5}. \nn
}
For the global phase we have
\eq{
\phi(\beta,c)\approx \phi_p=\arctan \frac{G_{odd}(\beta,c)}{H_{odd}(\beta,c)}. \label{eq.phi}
}
For the value $\beta=0.85$ which corresponds to the $\sqrt{s}=M_{\chi}$ we find  $\A_\chi\approx 5\cdot 10^{-4}$.
The quantity $D$ as function of  $y$, as defined in Eq. \ref{eq:eqy},
is  shown in Fig. \ref{fig.DEnergyPlot} for different values of $c$, $c=\cos(\vv{q_-}\vv{p_+})$. This quantity reaches its maximum value at the top of the resonance, where $y=0$. The angular dependence of the quantity $D$ is shown in Fig. \ref{fig.DAngularPlot} for different values of the total energy. One can see that $D$ is largest at backward and forward angle, and rapidly falls when deviating from the top of the resonance.

\begin{figure}
    \includegraphics[width=0.6\textwidth]{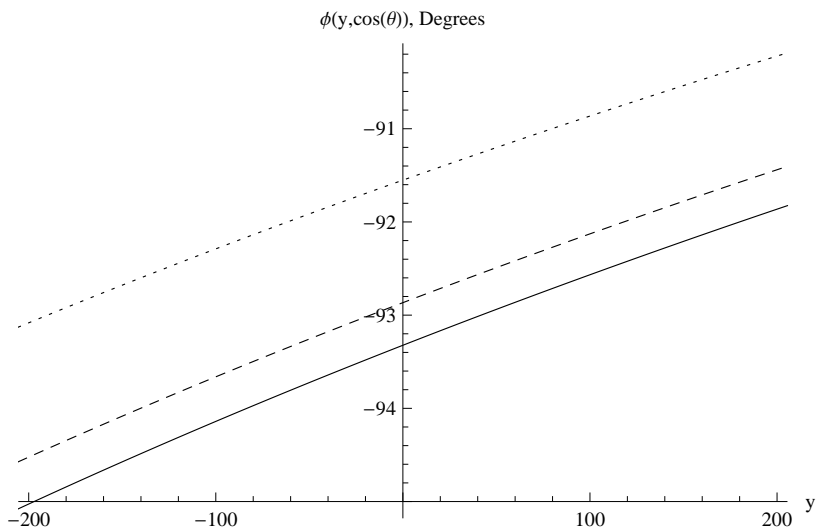}
    \caption{\label{fig.PhiEnergyPlot}
        The numerical estimation of the quantity $\phi(s)$ (in degrees) (see (\ref{eq.phi})) as a function of $y$ within the region corresponding to $M_\chi-100 \Gamma_\chi \leq \sqrt{s} \leq M_\chi+100\Gamma_\chi$.
        The solid line corresponds to $\theta=10^o$, dashed line --- $\theta=30^o$, dotted line --- $\theta=50^o$.
    }
\end{figure}

The total phase $\phi$, Eq. (\ref{eq.phi}) is of the order of $-90^\circ$ and shows a very weak dependence on the energy, see Fig.  \ref {fig.PhiEnergyPlot}.

\begin{figure}
    \includegraphics[width=0.6\textwidth]{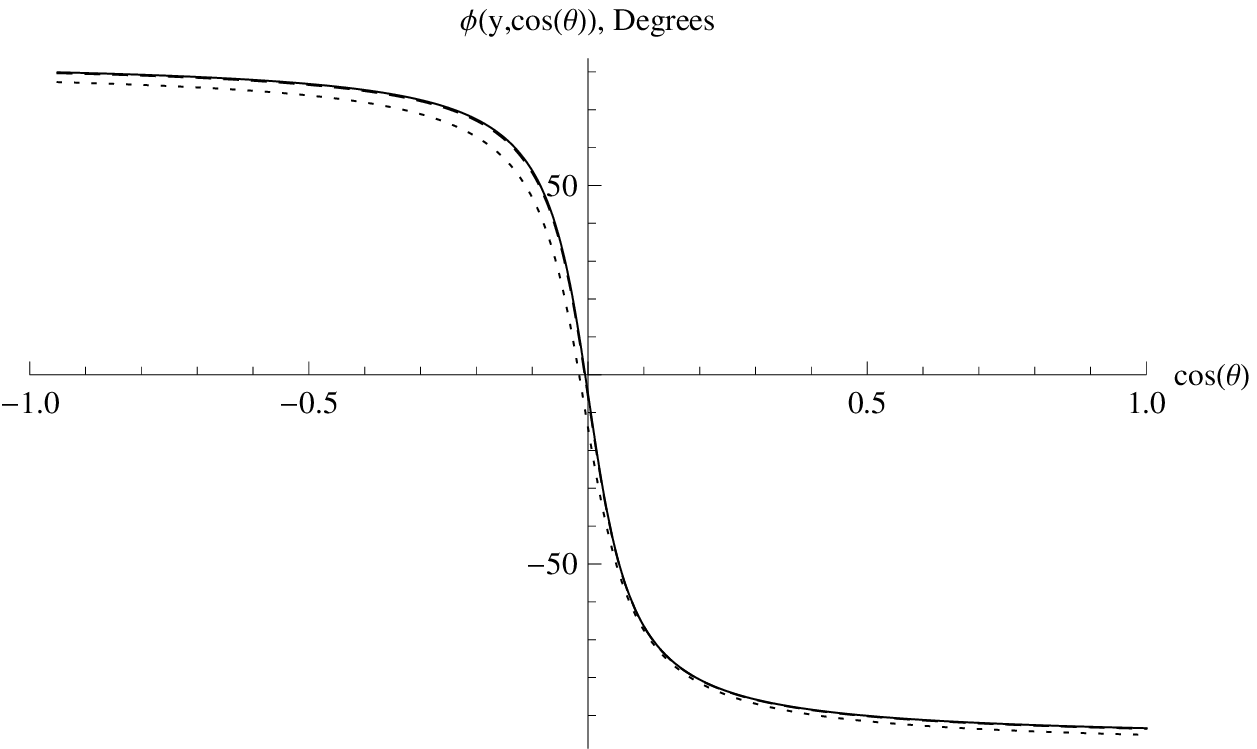}
    \caption{\label{fig.PhiAngularPlot}
        The numerical estimation of the quantity $\Psi(c)$ (see (\ref{eq.phi})) as a function of $c=\cos\theta$ for $\sqrt{s}=M_\chi$ and $F_1=1$, $F_2=0$.
        The solid line corresponds to $\sqrt{s}=M_\chi$, dashed line --- $\sqrt{s}=M_\chi-10\Gamma_\chi$, dotted line --- $\sqrt{s}=M_\chi-100\Gamma_\chi$.
    }
\end{figure}

The angular dependence of the total phase $\phi$, Eq. (\ref{eq.phi}), is  illustrated in Fig.  \ref {fig.PhiAngularPlot}.

\section{Z-boson and QED contributions to the asymmetry}

\begin{figure}
    \includegraphics[width=0.6\textwidth]{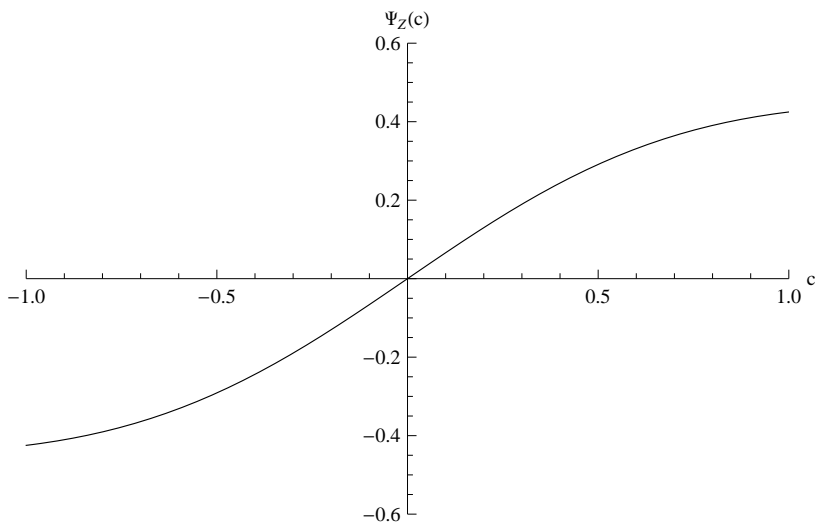}
    \caption{\label{fig.PsiZ} The numerical estimation of the quantity $\Psi_Z(c)$ (see (\ref{eq.PsiZ})) as a function of $c=\cos\theta$ for $\sqrt{s}=M_\chi$.}
\end{figure}

For comparison we calculate the contribution to the asymmetry arising from the interference of two Born amplitudes, with photon and
$Z$ boson intermediate states (we set here $F_1=1,F_2=0$):
\eq{
\A_Z=\A_0^Z \Psi_Z(c),
}
where
\eq{
\A_0^Z&=\frac{s}{M_Z^2\sin^2(2\theta_W)} \approx 2.1\cdot 10^{-3},\qquad \text{at $\sqrt{s}=M_\chi$},
\\
\Psi_Z(c)&=\frac{\beta c}{2-\beta^2(1-c^2)}, \label{eq.PsiZ}
}
with $\theta_W$ is the Weinberg angle, $\sin^2\theta_W\approx 0.231$. The  odd function $\Psi_Z(c)$ is shown in Fig.~\ref{fig.PsiZ}.
 One can see that this contribution, $\A_Z$ never exceeds $10^{-3}$.

\begin{figure}
    \includegraphics[width=0.6\textwidth]{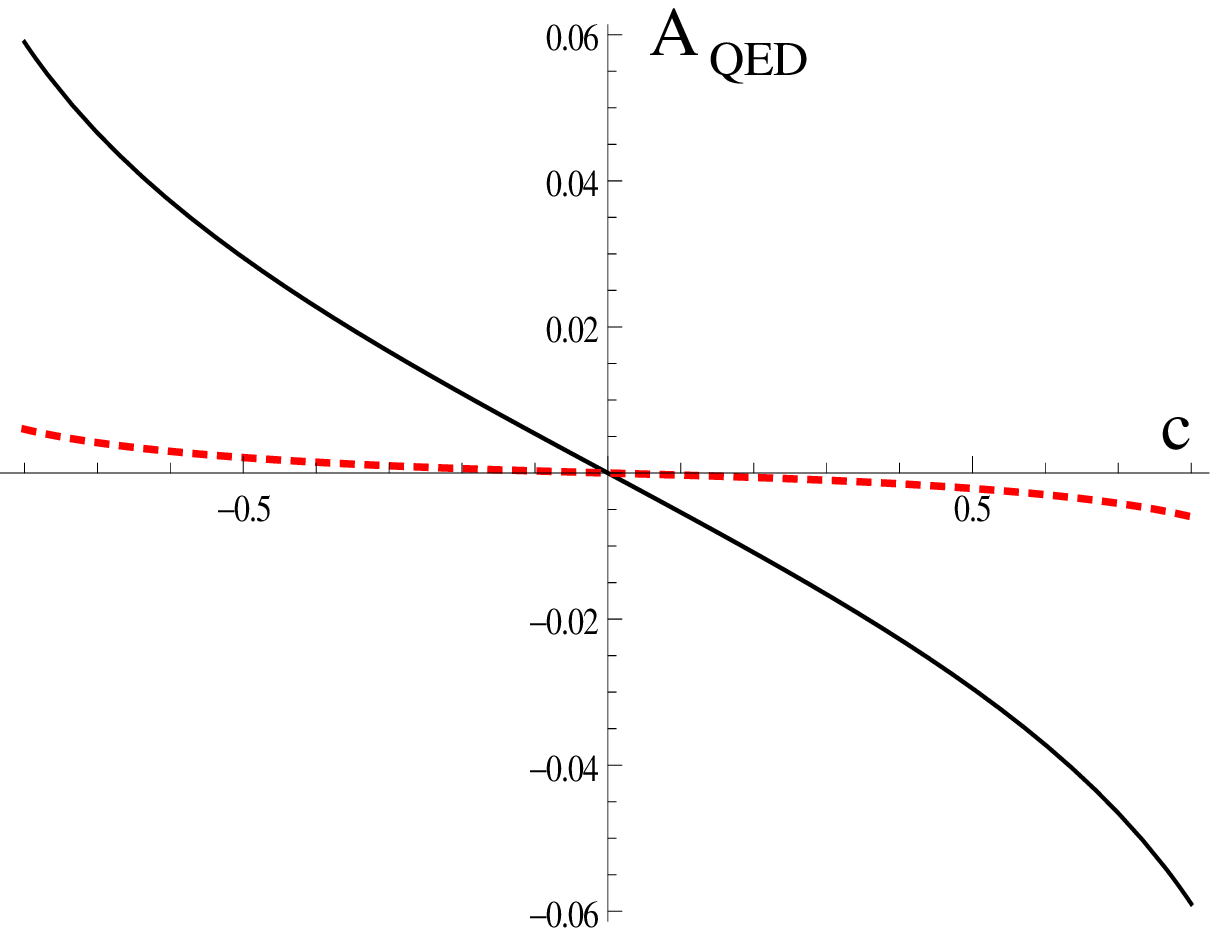}
    \caption{\label{fig.QED} Angular dependence of QED asymmetry in the $\chi_2$ resonance region: interference Born-box amplitudes (dashed line), soft and hard real photon emission (solid line).}
\end{figure}

Besides the two considered sources of the asymmetry, there is also a pure QED source, $\A_{QED}$. The total contribution to the odd part of the cross section was previously considered in Refs. \cite{KuraevMeledin,Ahmadov:2010ak}. The angular dependence of the asymmetry corresponding to this calculation, is plotted for $\sqrt{s}=M_{\chi}$ in Fig. \ref{fig.QED}. The dashed curve corresponds to the interference of the Born amplitude with the box ones, and it is very small. The dominant contribution arises from the soft and hard real photon emission from initial and final states (solid line).

\section{Conclusion}

We have calculated the charge asymmetry $\A$ for the reactions $p\bar p\leftrightarrow  e^+ e^-$, in the region of the $\chi_2$ resonance, due to different mechanisms: two photon exchange $\A_\chi$, Z-boson exchange, $\A_Z$, and the pure QED mechanism, $\A_{QED}$.

Our conclusion is that
\eq{
    10^{-2}  \sim |\A_{QED}| > |\A_Z| \geq |\A_\chi|  \sim 10^{-4}.
}

We have focussed to the energy region concerned by the formation of the $\chi_2$ resonance, because in a recent paper, Ref. \cite{Zhou:2011yz}, it was suggested that two photon exchange could make such asymmetry as large as 40\%, depending strongly on the relative phase between electric and magnetic form factors. Such phase was not fixed by the calculation and the results were given for any possible value of the phase.

Our results do not confirm such finding. We estimated this phase within a model which describes the $\chi_2\to p\bar p$ vertex with two gluon intermediate state which is assumed to saturate it. We show, indeed, that this asymmetry is maximum at the resonance, and at forward/backward angles, but its value keeps small ($\sim 10^{-4}$) due to the smallness of the constants. The coupling constants are expected to be small because such resonance is very narrow.

The possible reason of the discrepancy with Ref. \cite{Zhou:2011yz} can be related to the different description of the vertex
$\chi_2\to p\bar{p}$. In our case two gluons insure the strong interaction nature of this vertex, which is proportional to $\alpha_s$.

Let us note that the asymmetry measured in the process $p\bar p\to e^+ e^-$, as accessible at the PANDA (Darmstadt) is expected to be the same as in
$e^+ e^- \to p\bar p$, which can be measuured at BES (Beijing), as well as at VEPP (Novosibirsk) and DAFNE (Frascati).

\section{Acknowledgements}

Two of us (BVV), (EAK) are grateful to IPN Orsay for good working conditions during the solution of this problem.
EAK and YB are grateful to the grant RFBR № 11-02-00112 and BVV to the grant RFBR № 12-02-31703. for financial support.

\appendix
\section{The integrals for the amplitude of conversion of the $\chi_2(^3P_2)$ state into a lepton pair}
\label{appendix.A}

Let us derive the relevant integrals the amplitude of conversion of the 
$\chi_2(^3P_2)$ state into a lepton pair.

The phase volume of the $2\gamma$ intermediate state can be written as:
\eq{
\int \displaystyle\frac{d^4k_1}{(2\pi)^2}\delta(k_1^2)\delta(k_2^2)&=
\int  \displaystyle\frac{d^3k_1d^3k_2}{4\omega_1\omega_2(2\pi)^2}\delta^4(k_1+k_2-q)=\nn\\
&=\int  \displaystyle\frac{d\omega d \Omega}{16\pi^2} \delta(\sqrt{s}-2\omega)=
\displaystyle\frac{1}{16\pi}\int  \displaystyle\frac{d\Omega_1}{2\pi}.
}
Keeping in mind that $2\br{q_-k}=(s/2)(1-\beta c)$ and that $\int\frac{d \Omega}{2\pi}=\int\limits_{-1}^1 d c$, we define:
 \eq{
\int \displaystyle\frac{d \Omega}{2\pi}\displaystyle\frac{1}{2q_-k}&
\brf{ 1;k_{\mu};k_{\mu}k_{\nu};k_{\mu}k_{\nu}k_{\lambda}}=
\left\{I; \alpha q_{-\mu}+\beta q{\mu} \right. ;\nn\\
&
a_g g_{\mu\nu}+a_{--}q_{-\mu}q_{-\nu}+a_{qq}q_\mu q_\nu+ a_{q-}(q_\mu q_{-\nu}+q_{-\mu}q_\nu);~\nn\\
& a_{g-}( g_{\mu\nu}q_{-\lambda}+ g_{\mu\lambda}q_{\nu}+g_{\nu\lambda}q_{-\mu}) + a_{gq}( g_{\mu\nu}q_{\lambda}+g_{\mu \lambda}  q_{\nu}+ g_{\nu\nu} q_{\mu})
\nn\\
&+a_{---}q_{-\mu}q_{-\nu}q_{-\lambda}+ a_{qqq}q_{\mu}q_{\nu}q_{\lambda}+ a_{--q}(q_{-\mu}q_{-\nu}q_{\lambda}+q_{-\mu}q_{\nu}q_{-\lambda} +q_{\mu}q_{-\nu}q_{-\lambda})+\nn\\
&\left.a_{-qq}(q_{-\mu}q_{\nu}q_{\lambda}+q_{\mu}q_{-\nu}q_{\lambda} +q_{\mu}q_{\nu}q_{-\lambda})\right\}.
}
The calculation leads to:
\ga{
I=\displaystyle\frac{2}{s}\ln \displaystyle\frac{s}{m_e^2}; \quad
~ \alpha=I- \displaystyle\frac{4}{s}; \quad\beta= \displaystyle\frac{2}{s};\nn\\
a_g= -\displaystyle\frac{1}{2};~\quad a_{--}=I- \displaystyle\frac{6}{s}; \quad a_{qq}=a_{q-}= \displaystyle\frac{1}{s};
\nn\\
a_{g-}=\displaystyle\frac{2}{g}; ~\quad a_{gq}=-\displaystyle\frac{1}{3}; \quad a_{---}=I+\displaystyle\frac{4}{9s};\quad
a_{qqq}=\displaystyle\frac{4}{3s};~\quad a_{--q}=-\displaystyle\frac{2}{s};~\quad a_{-qq}=\displaystyle\frac{2}{3s}.
}
Using these expressions one obtains:
\eq{
\Delta_s\bar v(q_+) O_{\mu\nu} u(q_-)&=\displaystyle\frac{1}{16\pi }\sum_{spins}\int  \displaystyle\frac{d\Omega}{2\pi }\displaystyle\frac{1}{2\br{q_-k}}
\brs{\bar v(q_+)\dd{e_2}(\dd{q_-}-\dd{k}) \dd{e_1} u(q_-)}\times\nn\\
&\times\brs{(k_1k_2)e_1^\nu e_2^\mu +(e_1e_2)k_1^\nu k_2^\mu -k_1^\nu e_2^\mu (e_1 k_2)-k_2^\mu e_1^\nu (e_2k_1)}=\nn\\
&= \displaystyle\frac{1}{16\pi }\brs{\bar v(q_+)\gamma_\mu u(q_-)} q_{-\nu} \brs{s(I-2\alpha)- 4 a_{g_-} +sa_{--} +sa_{q-} +2a_g}=\nn\\
&= \displaystyle\frac{5}{72\pi }\brs{\bar v(q_+)\gamma_\mu u(q_-)}q_{-\nu}.
}
Performing the summation on the polarization states of $^3P_2$ quarkonium we obtain:
\eq{
&\sum_{\chi} \displaystyle\frac{1}{4} \Tr\brs{\dd{q_+} \gamma_\mu \dd{q_-} \gamma_\eta} \otimes
\displaystyle\frac{1}{4} \Tr\brs{\br{\dd{p_+} +M_p}\gamma_\lambda \br{\dd{p_-} -M_p}\Gamma_\eta}\chi_{\mu\nu} \chi_{\lambda\eta}=\nn\\
&=\beta c s^2M_p^2 \br{ -\frac{1}{4} }\br{ F_1+  \displaystyle\frac{s}{4M_p^2} F_2 } -  \displaystyle\frac{1}{16}(s\beta c) ^3 F_1.
}

\section{The angular integrals for proton tensor}
\label{appendix.B}

The angular integrals used for description of conversion of $^3P_2$ state to proton and antiproton are reported here:
\eq{
\int\frac{d\Omega_1}{2\pi(1-\beta c_1)} \brf{1;(1-c_2);(1-c_2)^2;(1-c_2)^3}=\brf{J_0;J_1;J_2;J_3}, \label{eq.AngularInegralsForProtonTensor}
}
where
\eq{
d\Omega_1=d c_1 \phi_1; \quad c_1=\cos(\vv{k_1}\vv{p_+}); \quad c_2=\cos(\vv{k_1}\vv{q_-})=c c_1-s s_1\cos\phi_1; \quad
c=\cos(\vv{q_-}\vv{p_+}). \nn
}
The result of integrations is
\eq{
J_0&=\frac{1}{\beta}L, L=\ln\frac{1+\beta}{1-\beta}; \nn \\
J_1&=\int\limits_{-1}^1\frac{d c_1(1-c c_1)}{1-\beta c_1};\nn \\
J_2&=\int\limits_{-1}^1\frac{d c_1[(1-c c_1)^2+(1-c^2)(1-c_1^2)/2]}{1-\beta c_1}; \nn \\
J_3&=\int\limits_{-1}^1\frac{d c_1[(1-c c_1)^2+3(1-c^2)(1-c_1^2)/2]}{1-\beta c_1}.
}
Let us note that the integrand in (\ref{eq.AngularInegralsForProtonTensor}) contains factor
$\br{1-\beta c_1}$, where $\beta$ is the proton velocity. This factor thus determines the lower limit of integration on $E_1^2$ in the dispersion integral used for real part of amplitude restoration (see (\ref{eq.DispersionRelation})):
\eq{
    H_{odd}(\beta,c) =
    \frac{\cal P}{2\pi}
    \int\limits_{M_p^2}^\infty \frac{dE_1^2}{E_1^2 - E^2}
    G_{odd}(E_1,c).
}


\end{document}